\documentclass[ reprint, amsmath,amssymb, aps,superscriptaddress,showpacs]{revtex4-1}

\usepackage{graphicx}
\usepackage{dcolumn}
\usepackage{bm}
\usepackage{xcolor}

\newcommand{\tens}[1]{\mbox{\textbf{\textit{\textsf{#1}}}}}

\usepackage[labelsep=endash]{caption}

\begin{document}

\preprint{APS/123-QED}

\title{Enhanced chiral discriminatory van der Waals interactions mediated by chiral surfaces
	}

\author{Pablo Barcellona}
\email{pablo.barcellona@physik.uni-freiburg.de}
\affiliation{Physikalisches Institut, Albert-Ludwigs-Universit\"at
Freiburg, Hermann-Herder-Str. 3, 79104 Freiburg, Germany}

\author{Hassan Safari}
\affiliation{Physics and Photonics Department, Graduate University of Advanced Technology, P.O. Box 76315-117, Mahan, Kerman, Iran} 

\author{A. Salam}
\email{salama@wfu.edu}
\affiliation{Department of Chemistry, Wake Forest University, Winston-Salem, NC 27109, USA} 

\author{Stefan Yoshi Buhmann}
\email{stefan.buhmann@physik.uni-freiburg.de}
\affiliation{Physikalisches Institut, Albert-Ludwigs-Universit\"at
Freiburg, Hermann-Herder-Str. 3, 79104 Freiburg, Germany}
\affiliation{Freiburg Institute for Advanced Studies,
Albert-Ludwigs-Universit\"at Freiburg, Albertstr. 19, 79104 Freiburg,
Germany}

\date{\today}

\begin{abstract}
We predict a discriminatory interaction between a chiral molecule and an achiral molecule which is mediated by a chiral body. To achieve this,
we generalize the van der Waals interaction potential between two ground-state molecules with electric, magnetic and chiral response to non-trivial environments. 
The force is evaluated using second-order perturbation theory with an effective Hamiltonian. Chiral media enhance or reduce the free interaction via many-body interactions, making it possible to measure the chiral contributions to the van der Waals force with current technology.
The van der Waals interaction is discriminatory with respect to enantiomers of different handedness and could be used to separate enantiomers.
We also suggest a specific geometric configuration where the electric contribution to the van der Waals interaction is zero, making the chiral component the dominant effect. 
\pacs{34.20.Cf, 33.55.+b, 33.80.-b,42.50.Nn}
\end{abstract}

\maketitle

\paragraph*{\label{Sec1}Introduction.}
Casimir and van der Waals (vdW) for\-ces are electromagnetic interactions
between neutral macro\-scopic bodies and/or molecules due to the  
quantum fluctuations of the electromagnetic field
\cite{Casimir48,CasimirPolder48,Milonni94}. In particular, the attractive vdW potential between two electrically polarisable particles was first derived by Casimir and Polder using the minimal-coupling Hamiltonian \cite{CasimirPolder48}. 
Molecules can also exhibit magnetic \cite{Mavroyannis,feinberg,feinberg2,safari2} and chiral polarisabilities \cite{Mavroyannis,jenkins,salam} and their contribution to the vdW force can be repulsive.

The aim of this work is the study of the interaction between chiral molecules in the presence of a chiral magneto-dielectric body.
Chiral molecules lack any center of inversion nor plane of symmetry. Hence they exist as two distinct enantiomers, left-handed and right-handed, which are related to space inversion. Due to their low symmetry they have distinctive interactions with light. In a chiral solution the refractive indices for circularly polarized light of different handedness are different. Hence a chiral solution can rotate the plane of polarization of light with an angle related to the concentration of the solution (optical rotation) \cite{Rosenfeld,Atkins,Powerchiral2}, or absorb left- and right-circularly polarised light at different rates  (circular dichroism) \cite{Powerchiral}. 
 All of these phenomena are related to the optical rotatory strength, defined in 
 terms of electric ($\textbf{d}_{nk}$) and magnetic ($\textbf{m}_{nk}$) dipole moment matrix elements \cite{craig}:
\begin{equation}
R_{nk} = \operatorname{Im} \left( {{{\mathbf{d}}_{nk}} \cdot {{\mathbf{m}}_{kn}}} \right)
.
\end{equation}
The rotatory strength changes  sign under spatial inversion.
Hence the contributions of the vdW force containing the rotatory strength are discriminatory with respect to the handedness of enantiomers \cite{jenkins,butcher,Buhmann}. According to the Curie dissymmetry principle \cite{curie}, the handedness of an enantiomer can only be detected by means of a second reference object which is also chiral (''It takes a thief to catch a thief''). This second object can be either a second chiral molecule,
or in our case, a chiral medium. 

Such discriminatory features could play an important role in the separation of enantiomers \cite{butcher}, which nowadays is  achieved with other methods like Coulomb explosion imaging \cite{pitzer}, rotational spectroscopy \cite{shubert}, and liquid chromatography \cite{kimata}. 

Previously, the vdW interaction between two chiral molecules was considered in free space \cite{jenkins}. In this scenario the chiral part of the vdW force is several orders of magnitude smaller than the purely electric contribution. We will show that this is no longer true when the chiral molecules are near a magneto-dielectric body which exhibits chiral properties.
Chiral media can be realized experimentally using tunable metamaterials \cite{soukoulis}, made from structures such
as a gold helix  \cite{gansel} or a twisted woodpile \cite{rill}, where large optical activity requires the lack of a mirror plane parallel to the substrate. These media will enhance or reduce the vdW potential via many-body interactions \cite{shajesh,milton}. Note that evidence for three-body dispersion forces has recently been found for the critical Casimir effect \cite{gambassi}.

The article is organized as follows.
We first develop an effective Hamiltonian in order to 
derive the vdW potential between two chiral molecules near magneto-dielectric bodies.
We then study the interaction between two chiral molecules in free space and subsequently introduce a chiral plate in order to enhance the chiral contribution. 

\paragraph*{\label{Sec2}Van der Waals potential.
}
In order to consider the vdW interaction of chiral molecules with a generic body, field quantization in an absorbing chiral medium is required \cite{butcher}. 
We introduce the vector field $\hat{\mathbf{f}}_\sigma\left( \mathbf{r},\omega  \right)$ 
as the bosonic annihilation
operator for the electric ($\sigma\!=\!0$) and magnetic ($\sigma\!=\!1$) excitations, respectively.  They obey the commutation
relation $
\left[  \hat{\mathbf{f}}_\sigma\left( \mathbf{r},\omega  \right),\hat{\mathbf{f}}_{\sigma '}^\dag \left( \mathbf{r}',\omega ' \right)\right] = \delta _{\sigma \sigma '}\bm{\delta} \left( \mathbf{r} - \mathbf{r}' \right)\delta \left( \omega  - \omega' \right)
$. Denoting by $|\{0\}\rangle$ the vacuum state of the electromagnetic field, $
\hat{\mathbf{f}}_\sigma \left( \mathbf{r},\omega  \right)|\{0\}\rangle=0 \quad \forall \,\sigma,\mathbf{r}, \omega\,,$
the excited photonic states are obtained by repeated action of the creation operator on the vacuum state. For instance, a two-photon excited state is given as
$\frac{1}{\sqrt{2}}\hat{\mathbf{f}}_{\sigma }^\dag( \mathbf{r},\omega )
\hat{\mathbf{f}}_{\sigma' }^{\dag}( \mathbf{r}',\omega'  )|\{0\}\rangle=
|\tens{1}_\sigma(\mathbf{r},\omega),\tens{1}_{\sigma'}(\mathbf{r}',\omega')\rangle.$
Introducing  the frequency component of the fields by $
\hat{\mathbf{E}}(\mathbf{r}) = \int_0^\infty \textrm{d}\omega \hat{\mathbf{E}}(\mathbf{r},\omega)+\text{H.c.}
$, which for the free field in the Heisenberg picture are simply Fourier components,
 we may write the electric field in terms of the bosonic annihilation operators as
\begin{equation}
 \hat {\mathbf{E}}\left( \mathbf{r},\omega  \right) = \int \text{d}^3 \mathbf{r'}  \sum\limits_\sigma  \tens{G}_\sigma \left( \mathbf{r},\mathbf{r'},\omega  \right) \cdot \hat {\mathbf{f}}_\sigma \left(\mathbf{r'},\omega  \right) 
\end{equation}
and $\hat {\mathbf{B}}\left( \mathbf{r},\omega  \right) = \nabla  \times \hat {\mathbf{E}}\left( \mathbf{r},\omega  \right)/\text{i}\omega $ with the tensors $\tens{G}_\sigma$ being defined in terms of the Green's tensor $\tens{G}$ according to Ref.~\cite{butcher}. The Green's tensor contains all geometrical as well as magneto-dielectric properties of the environment via the frequency-dependent relative permittivity, relative permeability and chiral susceptibility. It satisfies a useful integral relation with the mode-tensors $\tens{G}_\sigma$:
\begin{multline}
\sum\limits_{\sigma  =0,1} \int \text{d}^3\mathbf{s} \tens{G}_\sigma \left( \mathbf{r},\mathbf{s},\omega  \right) \cdot \tens{G}_\sigma ^{*\top} \left(\mathbf{r}', \mathbf{s},\omega  \right) =\\ \frac{\hbar \mu _0}{\pi }\omega ^2 \text{Im} \tens{G}\left( \mathbf{r},\mathbf{r}',\omega  \right),
\label{int-rel}
\end{multline}
where $\mu_0$ is the vacuum permeability.

The interaction between atoms (or molecules) and the field 
may be described 
by an effective interaction Hamiltonian \cite{passante}, which represents processes where two photons are created or annihilated at the same time:
\begin{equation}
\label{int-ham}
\hat H_{AF} =  - \frac{1}{2}\alpha _{ij}E_iE_j - \frac{1}{2}\chi _{ij}(E_iB_j -B_jE_i )
\end{equation}
where $\alpha$ is the electric polarizability and $\bm{\chi}=\bm{\chi}^{em}$ represents the chiral polarizability.
It can be expressed in terms of the electric and magnetic dipole moments: 
\begin{equation}
\bm{\chi}\left( \omega  \right) = \frac{1}{\hbar }\sum\limits_k \left( \frac{\mathbf{d}_{k0}\mathbf{m}_{0k}}{\omega _k^A + \omega } + \frac{\mathbf{d}_{0k}\mathbf{m}_{k0}}{\omega _k^A - \omega } \right),
\label{alpha}
\end{equation}
where $\omega _k^A = \left( {E_k^A - E_0^A} \right)/\hbar $ is the transition frequency between the excited state $k$ and the ground state.
In writing the effective Hamiltonian we have considered chiral molecules which respect time-reversal symmetry. In this case, Lloyd's theorem states that electric dipole moments are real, while magnetic dipole moments are purely imaginary \cite{lloyd}. We thus have $\bm{\chi}^{me}=-\bm{\chi}^{em \top}$. 
The chiral polarizability describes a mixed electric-magnetic response where an applied magnetic field contributes to an electric dipole moment in the atom:
\vspace{-0.1cm}\begin{align}\nonumber
\mathbf{\hat d} = &\bm{\alpha}  \left( \omega  \right)\cdot \hat {\mathbf{E}}\left( \omega  \right) + \bm{\chi}\left( \omega  \right) \cdot \hat {\mathbf{B}}\left( \omega  \right),\\
\mathbf{\hat m} =& -\bm{\chi}^{\top}\left( \omega  \right) \cdot \hat {\mathbf{E}}\left( \omega  \right). 
\end{align}
Thus the direction of the electric dipole moment is slightly rotated with respect to the direction of the electric field.
The chiral polarizability is discriminatory since it changes sign between different enantiomers. It vanishes for achiral molecules because of parity. 

\begin{figure}[ht]
   \centering
   \includegraphics[scale=0.51]{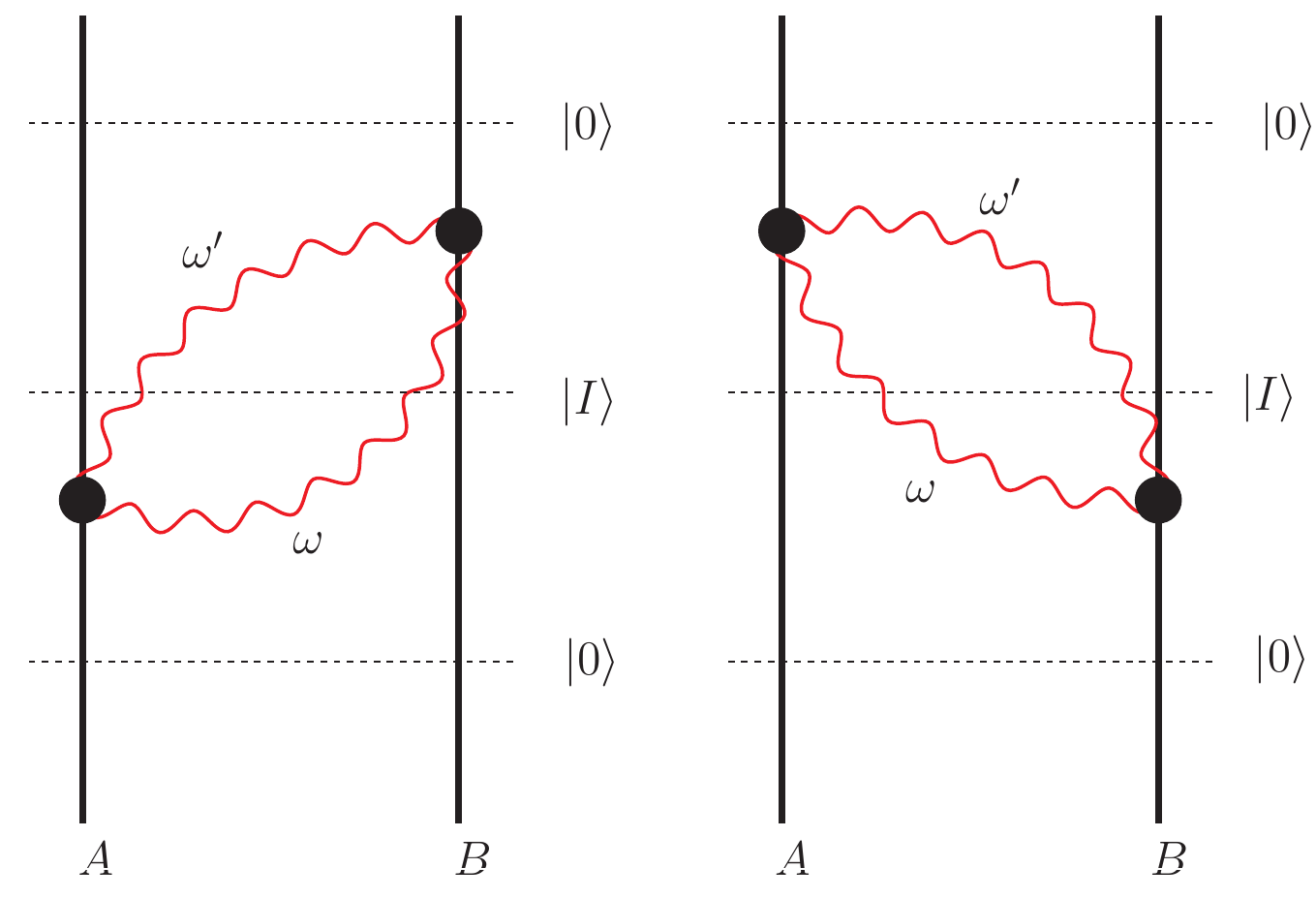}
   \caption{Van der Waals force between two molecules: two-photon exchange.}
    \label{fig1}
   \end{figure}

With the interaction Hamiltonian being quadratic in the fundamental operators $\hat{\mathbf{f}}_\lambda$ and $\hat{\mathbf{f}}_\lambda^\dagger$, it is seen that the energy shift describing the interaction between the two molecules A and B can be obtained from 
second order perturbation theory 
\begin{equation}
\label{shift}
\Delta E_{(2)} = -\sum_{I}\frac{\langle 0|\hat{H}_\textrm{int}|I\rangle\langle I|\hat{H}_\textrm{int}|0\rangle}{E_I-E_0}\equiv U(\mathbf{r}_A,\mathbf{r}_B),
\end{equation} 
where $|0\rangle \!\equiv\!|0\rangle_A|0\rangle_B|\{0\}\rangle$ and $\hat{H}_\textrm{int}\!=\!\hat{H}_{AF}+\hat{H}_{BF}$. The intermediate state $|I\rangle$ corresponds to a state in which both molecules are in their ground state with two virtual photons being present. The relevant two-photon processes are represented by the Feynman diagrams in Fig.~\ref{fig1}.
   The collective sum in Eq.~(\ref{shift}) runs over both discrete and continuous parameters. Using the interaction Hamiltonian (\ref{int-ham}), making use of the integral relation (\ref{int-rel}), and simplifying the resulting expression using contour integral technique in the complex frequency plane,
the purely electric and electric-chiral energy shifts take the final forms 
\begin{subequations}
\begin{align}\nonumber
U^{EE}\left( \textbf{r}_A,\textbf{r}_B \right) =&  - \frac{\hbar \mu _0^2}{2\pi }\int\limits_0^\infty  \text{d}\xi  \xi ^4 \text{Tr}\Big\{ \bm{\alpha} _A\left( \text{i}\xi  \right) \\
&\cdot \tens{G}\left( \textbf{r}_A,\textbf{r}_B,\text{i}\xi  \right) \cdot \bm{\alpha} _B\left( \text{i}\xi  \right)\cdot \tens{G}\left( \textbf{r}_B,\textbf{r}_A,\text{i}\xi \right) \Big\}\\ \nonumber
U^{CE}\left( \mathbf{r}_A,\mathbf{r}_B \right) =& \frac{\hbar \mu _0^2}{\pi }\int\limits_0^\infty  \text{d}\xi  \xi ^3\text{Tr}\Big\{ \bm{\chi} _A\left( \text{i}\xi  \right)\\
&\hspace{-2cm}\cdot \nabla _A \times \tens{G}\left( \mathbf{r}_A,\mathbf{r}_B,\text{i}\xi  \right)\cdot \bm{\alpha} _B\left( \text{i}\xi  \right)\cdot \tens{G}\left( \mathbf{r}_B,\mathbf{r}_A,\text{i}\xi  \right) \Big\}.
\end{align}
\end{subequations}
 Resonant terms in the molecular frequencies have been avoided by means of infinitesimal imaginary shifts.
The purely electric (achiral) contribution is well known in literature \cite{safari2,safari3}.
Other terms contributing to the total van der Waals interaction are given in the supplementary material.

\paragraph*{\label{Sec3}Chiral molecules in free space.}
 To apply our general results to free space, one uses the bulk Green's tensor
 \begin{equation}
\tens{G}^{(0)}\left( \mathbf{r}_A,\mathbf{r}_B,\text{i}\xi  \right) = \frac{c^2\text{e}^{ - \xi r/c}}{4\pi \xi ^2r^3}\left[ a\left( \frac{\xi r}{c} \right)\tens{I} - b\left( \frac{\xi r}{c} \right){\mathbf{e}_r} \mathbf{e}_r \right],
 \end{equation}
where $c$ is the speed of light, $\textbf{r}=\textbf{r}_B-\textbf{r}_A$, $r=|\textbf{r}|$, $\mathbf{e}_r=\textbf{r}/r$, $a\left( x \right) = 1 + x + x^2$, and $ b\left( x \right) = 3 + 3x + x^2$.
 
The order of magnitude of a specific contribution of the vdW energy is roughly given by the respective polarizability, and thus depends on the numbers of electric and magnetic dipole moments. 
An electric dipole has an order of magnitude of $e a_B$, where $a_B$ is the Bohr radius, while the magnetic moment has an order of magnitude of the Bohr magneton  $e \hbar /2m$. Hence $\frac{{m/c}}{d}$ is of the order of the fine-structure constant. 
The leading order contribution is that from the purely electric term, which contains four electric dipole moments. 
 
  The next-to-leading term is the  
chiral-electric contribution, $U^{CE}\left( r \right)$, which contains three electric and one magnetic dipole moments.
  However, if the chiral polarizability of the molecule is isotropic, $U^{CE}\left(r \right)=0$.
  This vanishing interaction is a consequence of the Curie dissymmetry principle \cite{curie}: the vdW potential cannot distinguish the handedness of the chiral molecule if the other molecule is not chiral.

The next-order terms contain two electric and two magnetic dipole moments: $U^{EM}\left( r \right) $ and $U^{CC}\left( r \right) $.
For isotropic polarizabilities, it has been shown that \cite{jenkins,salam}:
 \begin{equation}
U^{CC}\left( r \right) =   \frac{\hbar }{8\pi ^3\varepsilon _0^2r^6}\int\limits_0^\infty  \text{d}\xi \frac{\chi _A\left( \text{i}\xi  \right)}{c}\frac{\chi _B\left( \text{i}\xi  \right)}{c} l\left( \frac{\xi r}{c} \right), \\
 \end{equation}
 where $l\left( x \right) = \text{e}^{ - 2x}\left( 3 + 6x + 4x^2 \right)$ and $\epsilon_0$ is the vacuum permittivity. The interaction scales as $r^{-6}$ in the non-retarded limit, but decreases more rapidly in the far-zone ($r^{-9}$) due to the finite velocity of  
light. This result can be obtained in our formalism by using the vacuum Green's tensor.
 
 Because the purely electric energy contains four electric dipole moments and $U^{CC}$ contains two electric and two magnetic dipole moments, the latter is roughly four orders of magnitude smaller than the former. This shows the necessity to use magneto-dielectric bodies to make the chiral contribution appreciable.
 While $U^{CE}\left( r \right)$ is strictly zero in free space for isotropic molecules, this is no longer true with general material environments, which exhibit chiral properties.

\paragraph*{\label{Sec4}Chiral molecules near a perfect chiral plate.}
 
\begin{figure}[htbp]
   \centering
   \includegraphics[scale=0.60]{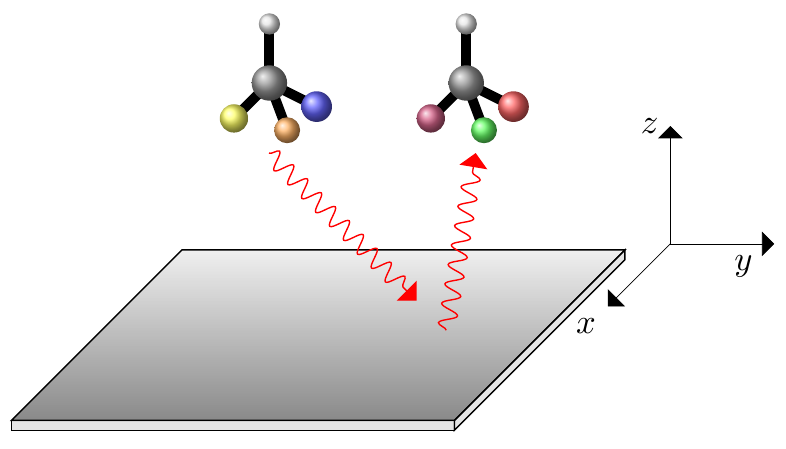}
   \caption{medium-assisted vdW interaction between two molecules: direct or indirect exchange of two photons. }
   \label{fig2}
   \end{figure}

With the introduction of a body, the fluctuations of the electromagnetic field are influenced by the presence of the boundary.    
The interaction between chiral molecules is a two-photon exchange, where photons can be reflected by the body's surface (Fig. \ref{fig2}). 
  The total Green's tensor is the sum of the bulk (free-space) Green's tensor $\tens{G}^{(0)}$ and the scattering Green's tensor of the plate $\tens{G}^{(1)}$, that accounts for the reflection of the electromagnetic field from surfaces:
$ \tens{G} =\tens{G}^{(0)} +\tens{G}^{(1)}$.

If one species is chiral and the other one is achiral we need a chiral surface to observe a discriminatory effect. 
 The simplest surface is a perfect chiral plate which reflects a $p$-polarized wave to an $s$-polarized wave and \textit{vice versa} with reflection coefficients $r_{s \to p}$ and $r_{p \to s}$. This is a limit of a chiral plate, which may approached for metamaterials with large optical activity. In other words, it rotates the polarisation of an incoming wave by an angle of $\pm \pi/2$ when looking along the direction of the wave vector. Its scattering Green's tensor reads:
\begin{multline}
\tens{G}^{(1)}\left( \mathbf{r}_A,\mathbf{r}_B,\text{i}\xi  \right) = \frac{1}{8\pi ^2}\int\limits_0^{2\pi } \text{d}\varphi  \int\limits_0^\infty  \text{d} k^\parallel \frac{k^\parallel }{\kappa ^ \bot }\text{e}^{ - \kappa ^ \bot z_ + }\\ 
 \hspace{-0.2cm}  \times \text{e}^{-\text{i}k^\parallel \left( x\cos \varphi  + y\sin \varphi  \right)} \left( \mathbf{e}_{p + }\mathbf{e}_{s - }r_{s \to p} + \mathbf{e}_{s + }\mathbf{e}_{p - }r_{p \to s} \right),
 \end{multline}
 where $x=x_B-x_A$, $y=y_B-y_A$, $ z_+=z_A+z_B$ and we have introduced the polarization unit vectors $\mathbf{e}_{s \pm } = \left( \sin \varphi , - \cos \varphi ,0 \right)$, $\mathbf{e}_{p \pm } = \frac{c}{\xi }\left(  \mp \kappa ^ \bot \cos \varphi , \mp \kappa ^ \bot \sin \varphi , - \text{i}k^\parallel  \right)$.
The parallel and perpendicular components of the wave vector, $k^\parallel$ and $ \kappa^\bot$, are related via ${\kappa ^ \bot } = \sqrt {\frac{{{\xi ^2}}}{{{c^2}}} + {k^{\parallel 2}}} $. 
For a plate of posi\-tive chirality $r_{s \to p}=-1$, $r_{p \to s}=1$, and for a plate of negative chirality $r_{s \to p}=1$, $r_{p \to s}=-1$.

The perfect chiral plate has no influence on the purely electric contribution, which reduces to the well-known electric van der Waals interaction in free space.
Instead, it influences the leading-order chiral contribution, which in the non-retarded limit (all distances much lower than the relevant atomic transition wavelength) reads:
\begin{multline}
U_{\text{nr}}^{CE} =  \pm \frac{\hbar }{16\pi ^3\varepsilon _0^2} \int\limits_0^\infty  \text{d}\xi \frac{\chi _A\left( \text{i}\xi  \right)}{c}\alpha _B\left( \text{i}\xi  \right)
\label{UCE} \\ 
\times \frac{r^2\left[ 2r_ + ^2 - 3\left( x^2 + y^2 \right) \right] - 3r_ + ^2\left( x^2 + y^2 \right)}{r^5r_ + ^5}
\end{multline}
where $r_ +  = \sqrt {x^2 + y^2 + z_ + ^2}$. If the two species are chiral, there is another discriminatory term $U^{CC}$ which is however much smaller than $U^{CE}$.

Consider, for example, the interaction between a Rubidium atom and the chiral molecule 3-methylcyclopentanone (3-MCP) \cite{steck,kroner}. The molecule 3-MCP has a particularly small electric dipole moment and a magnetic dipole moment comparable to the Bohr magneton; this kind of system is suitable to observe enhanced chiral effects.

We suppose the position of the molecule A is fixed and we write the resulting energy 
in terms  of the interparticle separation, $\textbf{r}=\textbf{r}_B-\textbf{r}_A$; 
$U\left( \mathbf{r}_A,\mathbf{r}_A + \mathbf{r} \right)$. The total force acting on B is $\textbf{F}=-\nabla _{\mathbf{r}}
U\left( \mathbf{r}_A,\mathbf{r}_A + \mathbf{r} \right)$ and its direction differs from $\mathbf{r}$ because of the presence of the additional surface. 
We choose, without loss of generality, a coordinate system such that the plate corresponds to the $z=0$ plane, both species lie in the $x=0$ plane with the coordinates of molecule $A$
being $\mathbf{r}_A = \left( 0,0,z_A \right)$. 
The degree of  attractiveness is quantified by using the parameter $\mathbf{e}_r \cdot \mathbf{F}$, {where $\mathbf{e}_r =\textbf{r}/r$.

 A three-dimensional plot shows the relative importance between electric and chiral contributions for a plate of negative chirality (see Fig.~\ref{fig3}).
 \begin{figure}[htbp]
   \centering
   \includegraphics[scale=0.35]{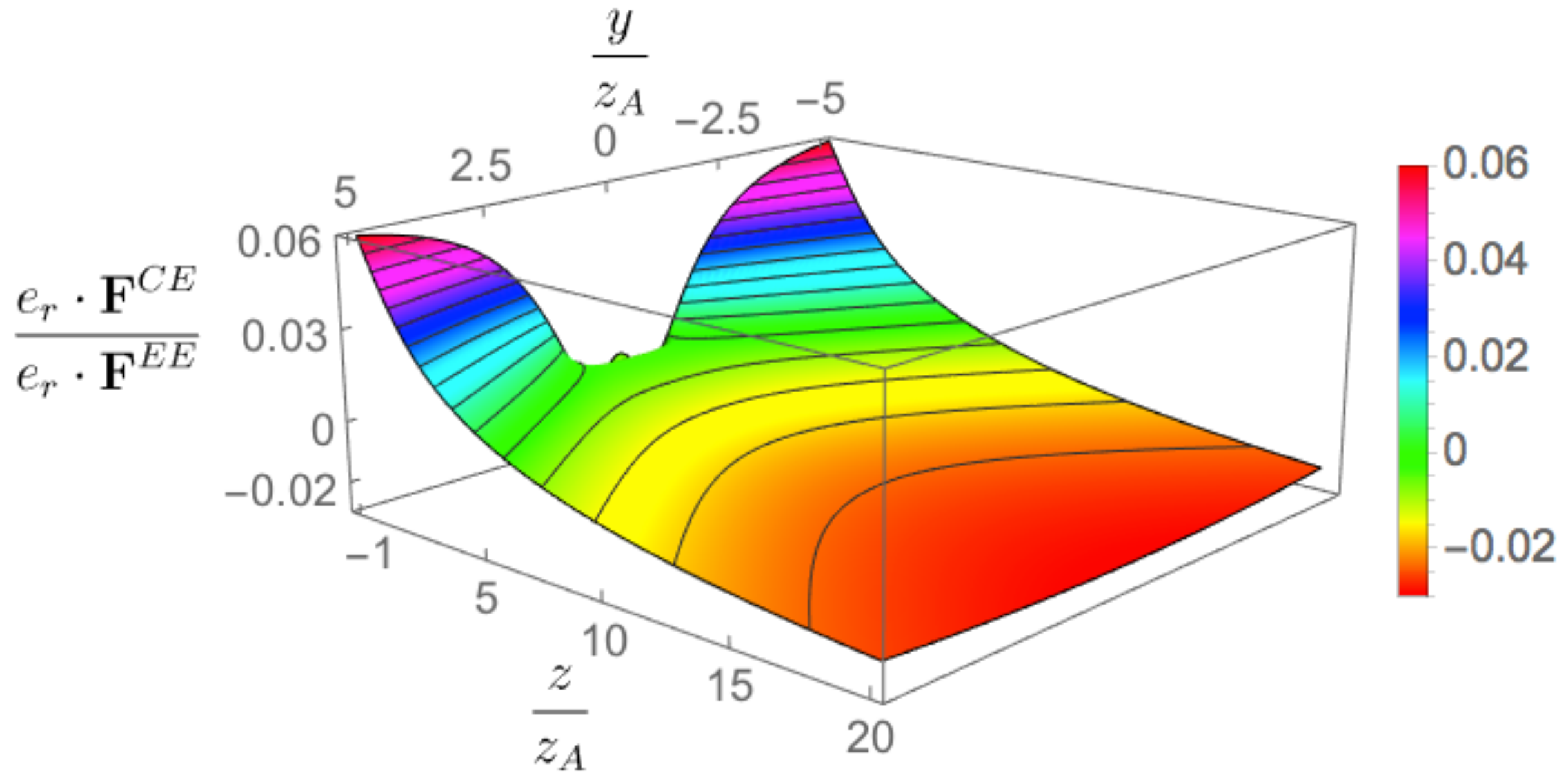}
   \caption{$\mathbf{e}_r \cdot \mathbf{F}^{CE}/\mathbf{e}_r \cdot \mathbf{F}^{EE}$ for a Rubidium atom (atom B) and the chiral molecule 3-MCP (molecule A). }
   \label{fig3}
   \end{figure}
     The chiral contribution is maximally enhanced when both species are aligned parallel to the surface ($z=0$), having a large distance along the $y$-direction ($y/z_A \gg 1$). In this case the chiral contribution is attractive and one order of magnitude smaller than the purely electric contribution ($\mathbf{e}_r \cdot \mathbf{F}^{CE}/ \mathbf{e}_r\cdot \mathbf{F}^{EE} \simeq 6.75 \% $). 
When both species are aligned perpendicular to the surface ($y=0$) and have a large distance along the $z$-direction ($z/z_A \gg 1$), the chiral contribution is repulsive
   and two orders of magnitude smaller than the electric contribution ($\mathbf{e}_r \cdot \mathbf{F}^{CE}/\mathbf{e}_r\cdot \mathbf{F}^{EE} \simeq -3.37 \%$).
    
Our chiral contribution could be observed in Rydberg molecules, which have large dipole moments and therefore strong van der Waals interactions \cite{crosse}.
  We expect a strongly enhanced interaction between a Rydberg molecule and a chiral molecule, which does not have to be in a Rydberg state.
On the other hand it has been recognized that it is possible to create chiral Rydberg molecules, exciting moving Rydberg atoms of Rubidium by two right- or two left-circularly polarized photons \cite{hammer}. Such systems could be used in a proof-of-principle experiment. If two identical chiral Rydberg molecules are exposed to a resonant laser, the laser can cause the system to oscillate between the ground state and the excited states. From the oscillations of the molecular populations  it is possible to deduce the van der Waals interaction. The experiment has been performed with Rubidium atoms where 
$C_6$ (van der Waals coefficient of the energy $U=-C_6/r^6$) has been determined with an accuracy of $2.7\% $ for the excited Rydberg state $n=62$ \cite{beguin}.
Note that if one molecule is absorbed onto the surface the strength of the interaction is even higher because of the influence of the substrate on the molecular polarizability, as shown in Ref. \cite{kawai}.


Most importantly, the chiral contribution is discriminatory: it changes sign if we substitute a chiral molecule with its enantiomer or a positive chiral plate with a negative one. Hence by determining whether the free interaction is enhanced or reduced by the presence of the plate, it could be possible to identify the handedness of the chiral molecule.
This discriminatory effect also suggests  configurations, where the non-chiral contributions of the van der Waals force completely vanish, but not the chiral contributions.  Fig. \ref{fig4} shows such possible geometry.
 \begin{figure}[htbp]
   \centering
   \includegraphics[scale=0.70]{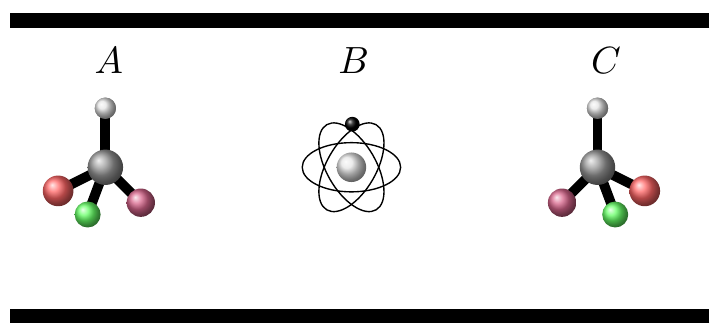}
   \caption{Three species parallel in the symmetry axis of a cavity. }
   \label{fig4}
   \end{figure}
  Three species are in the symmetry axis of a cavity composed of two identical chiral plates. We are interested in the force acting on the middle achiral atom $B$.  
This experiment allows us to establish if the two chiral molecules $A$ and $C$ have the same handedness. If $A$ and $C$ have the same handedness
then the force acting on $B$ is strictly zero; this is no longer true if they are opposite enantiomers
 because of chiral discriminatory contributions of the vdW force.
This interaction is enantiomer selective because  
atom $B$ will be attracted or repelled from 
 molecule A depending on the handedness of the latter. Hence if we know the handedness of one chiral molecule we are able to determine the handedness of the other one with this experiment.
  Note that the recently predicted exponential suppression of the electric vdW potential in a parallel mirror cavity \cite{melo} or in a planar waveguide \cite{haak} might further assist the observability of the chiral component. 
Three-body discriminatory chiral contributions are also present in the cavity configuration, however they are smaller with respect to the analyzed two-body contributions.
 
Alternatively, in scattering experiments one commonly has molecular beams of chiral molecules containing both
enantiomers  in equal proportions. This molecular beam can travel near a gas of chiral molecules of one handedness, prepared for example with liquid chromatography. The interactions with the gas molecules will lead to a separation of the trajectories of the molecules in the beam, depending on their handedness. This trajectory imaging technique has been already used for Rubidium Rydberg atoms \cite{thaicharoen}.

\paragraph*{\label{Sec5}Conclusions.}
Our model, which combines time-inde\-pendent perturbation theory with an effective-Hamil\-tonian approach, has allowed us to study the influence of material environments on the vdW interaction involving chiral molecules.

We see that surfaces with chiral properties can significantly enhance chiral contributions. The chiral contribution can be repulsive unlike the purely electric force which is always attractive. For some positions the plate enhances the free interaction (attractive chiral contribution) and for other positions it reduces the interaction (repulsive chiral contribution).

 Secondly, the  chiral force is discriminatory with respect to enantiomers of different handedness, thus opening an interesting perspective on the separation of enantiomers. We suggest symmetric configurations where we selectively cancel the purely electric contribution, but not the chiral components:
this selective cancellation is a direct consequence of the  non-discriminatory nature of electric and magnetic interactions, and the
 discriminatory nature of the chiral components.

\paragraph*{Acknowledgement.--}
This work was supported
by the Deutsche Forschungsgemeinschaft (Grants BU 1803/3-1 and GRK 2079/1) and the Freiburg
Institute for Advanced Studies (FRIAS).

\subparagraph*{Supplementary material}
\paragraph*{Van der Waals potential in absorbing media.
}
In order to consider the vdW interaction of chiral molecules with a generic body, field quantization in an absorbing chiral medium is required \cite{butcher}. 
In order to achieve a compact notation, let us represent electric and magnetic quantities, by numeric labels 0 and 1, respectively.

For example, we introduce dual electromagnetic fields as $\hat{\mathbf{F}}_0 \!=\! \hat{\mathbf{E}}$ and $\hat{\mathbf{F}}_1\! =\! c\hat{\mathbf{B}}$, and the frequency component of the fields by
\begin{equation}
\label{fre-comp}
\hat{\mathbf{F}}_\lambda(\mathbf{r}) = \int_0^\infty \textrm{d}\omega \hat{\mathbf{F}}_\lambda(\mathbf{r},\omega)+\text{H.c.}\,,
\quad \lambda=0,1
\end{equation}
 we may write the fields in terms of the bosonic annihilation operators $\textbf{f}_\sigma$ as
\begin{equation}
\label{field}
\hat{\mathbf{F}}_{\lambda}\left( \mathbf{r},\omega  \right) =  - \frac{c^2}{\omega ^2}\sum\limits_{\sigma = 0,1} \int \text{d}^3 \mathbf{r'} \tens{G}_{\sigma}^{\lambda 0} \left( \mathbf{r},\mathbf{r'},\omega  \right) \cdot \hat{\mathbf{f}}_{\sigma} \left( \mathbf{r'},\omega  \right),
\end{equation}
with the tensors $\tens{G}_\sigma$ being defined in terms of the Green's tensor $\tens{G}$ according to Ref.~\cite{butcher}. They satisfiy the useful integral relation with the mode-tensors $\tens{G}_\sigma$:
\begin{multline}
\sum\limits_{\sigma = 0,1} \int \text{d}^3 \mathbf{s}\left[ \tens{G}_\sigma \left( \mathbf{r},\mathbf{s},\omega  \right) \right]^{\lambda 0} \cdot \left[ \tens{G}_\sigma ^*\left( \mathbf{s},\mathbf{r'},\omega  \right) \right]^{0\lambda'} \\
=  - \frac{\hbar \mu _0}{\pi c^2}\omega ^4\left[ \text{Im} \tens{G}\left( \mathbf{r},\mathbf{r'},\omega  \right) \right]^{\lambda \lambda'}.
\label{int-rel}
\end{multline}
 In addition, the superscripts  $\lambda\lambda'$ used in Eq.~(\ref{field}) and Eq.~(\ref{int-rel}), for an arbitrary tensor 
$\tens{T}( \mathbf{r},\mathbf{r'},\omega)$ are defined as follows:
\begin{align}
	\label{tensors}
\tens{T}^{00}&=  - \frac{\omega ^2}{c^2} \tens{T},&\tens{T}^{01}	&= \frac{\text{i} \omega }{c} \tens{T} \times \overleftarrow \nabla',
&\nonumber\\
\tens{T}^{10}&= \frac{\text{i} \omega }{c}  \overrightarrow \nabla \times   \tens{T}, &
\tens{T}^{11}&=\overrightarrow \nabla   \times \tens{T} \times \overleftarrow \nabla '.&
	\end{align}

The interaction between atoms (molecules) and field 
may be described 
by an effective interaction Hamiltonian \cite{passante}, 
\begin{multline}
\label{int-ham}
\hat{H}_{AF} =  - \frac{1}{2}\sum_{\lambda,\lambda'}\int\limits_0^\infty  {\text{d}} \omega \hat{\mathbf{F}}_{{\lambda}}\left( {\mathbf{r}_A} \right) \cdot \bm{\alpha} _A^{\lambda\lambda'}\left( \omega  \right) \cdot \hat{\mathbf{F}}_{\lambda'}\left( \mathbf{r}_A,\omega  \right)  \\+ \text{H.c.}
\end{multline}
where 
$\bm{\alpha}^{\lambda\lambda'}$  
represents the ground-state electric ($\lambda\!=\!\lambda'\!=\!0$), magnetic
($\lambda\!=\!\lambda'\!=\!1$) or chiral ($\lambda\!\neq\!\lambda'$)  polarizability of the molecule. 
With the assignments 
$\hat{\bm{\mu}}^0 = \hat{\mathbf{d}}$, $\hat{\bm{\mu}}^1 =\hat{ \mathbf{m}}/c$  
the polarizabilities 
can be written in 
a general form:
\begin{equation}
\bm{\alpha}_A^{\lambda\lambda'}\left( \omega  \right) = \frac{1}{\hbar } \sum_k  
 \left( \frac{\bm{\mu}_{k0}^\lambda
\bm{\mu}_{0k}^{\lambda'}}{ \omega _k^A + \omega  }  + \frac{\bm{\mu}_{0k}^\lambda
\bm{\mu}_{k0}^{\lambda'}}{\omega _k^A - \omega   } \right),
\label{alpha}
\end{equation}
where $\omega _k^A = \left( {E_k^A - E_0^A} \right)/\hbar $ is the transition frequency between the excited state $k$ and the ground state.
We will consider time-reversal symmetric systems, for which the electric dipole elements are real and the magnetic dipole elements are purely immaginary. 
Note that the Hamiltonian, in Eq.  \ref{int-ham}, generalizes the effective Hamiltonian in the main text for dynamical polarizabilities.

The energy shift describing the interaction between the two molecules A and B can be obtained from 
second order perturbation theory. The two-photon matrix element reads:
\begin{multline}\left\langle 0 \right|H_{AF}\left| \mathbf{1}_\sigma \left( \mathbf{r},\omega  \right),\mathbf{1}_{\sigma '}\left( \mathbf{r}',\omega ' \right) \right\rangle  =\\
 =  - \frac{1}{2\sqrt 2 }\frac{c^4}{\omega ^2 \omega '^2}\sum\limits_{\lambda ,\lambda'}\left(  - 1 \right)^{\lambda'} \left[ \tens{G}_\sigma \left( \mathbf{r},\mathbf{r}_A,\omega  \right) \right]^{0\lambda}
 \\ \cdot  \left[ \bm{\alpha} _A^{\lambda \lambda'}\left( \omega  \right) + \bm{\alpha}  _A^{\lambda \lambda'}\left( -\omega ' \right) \right]\cdot  \left[ \tens{G}_{\sigma '}\left( \mathbf{r}_A,\mathbf{r'},\omega ' \right) \right]^{\lambda'0}
\label{1}\end{multline}
Using the integral relation Eq. (\ref{int-rel}) and the transpose properties of the Green's tensors and polarizabilities we find: 
\begin{multline}
U\left( \mathbf{r}_A,\mathbf{r}_B \right) =\\
  - \frac{\hbar }{2\pi ^2\varepsilon _0^2} \sum\limits_{\lambda _1,\lambda _2,\lambda _3,\lambda _4}  \int\limits_0^\infty  \text{d} \omega \int\limits_0^\infty  \text{d} \omega '\frac{1}{\omega  + \omega '}\left(  - 1 \right)^{\lambda _2 + \lambda _4}\\ \times \text{Tr}\left\{ \frac{\bm{\alpha} _A^{\lambda _1\lambda _2}\left( \omega  \right) +\bm{\alpha}_A^{\lambda _1\lambda _2}\left( -\omega ' \right)}{2} \cdot \left[ \text{Im}\tens{G}\left( \mathbf{r}_A,\mathbf{r}_B,\omega ' \right) \right]^{\lambda _2\lambda _4} \right.\\
 \cdot \left.\frac{\bm{\alpha} _B^{\lambda _4\lambda _3}\left( \omega  \right) +\bm{\alpha} _B^{\lambda _4\lambda _3}\left(- \omega ' \right)}{2} \cdot \left[ \text{Im} \tens{G}\left( \mathbf{r}_B,\mathbf{r}_A,\omega  \right) \right]^{\lambda _3\lambda _1} \right\} \\
 + \left( A \leftrightarrow B \right).
 \label{2}
\end{multline}
The rotation of the integral in the imaginary axis gives:
\begin{eqnarray}
&&U\left( \mathbf{r}_A,\mathbf{r}_B \right) =\sum\limits_{\lambda _1,\lambda _2,\lambda _3,\lambda _4} U_{\lambda _1\lambda _2\lambda _3\lambda _4}\left( \mathbf{r}_A,\mathbf{r}_B \right), \\
&&U_{\lambda _1\lambda _2\lambda _3\lambda _4}\left( \mathbf{r}_A,\mathbf{r}_B \right) =- \frac{\hbar }{2\pi \varepsilon _0^2} \int\limits_0^\infty  \text{d} \xi \text{Tr}\left[ \bm{\alpha} _A^{\lambda _1\lambda _2}\left( \text{i}\xi  \right) \right.\nonumber\\
&& \hspace{-2ex}\left.  \cdot  \tens{G}^{\lambda _2\lambda _3}\left( \mathbf{r}_A,\mathbf{r}_B,\text{i}\xi  \right)  
\cdot\bm{\alpha}_B^{\lambda _3\lambda _4}\left( \text{i}\xi  \right)\cdot\tens{G}^{\lambda_4\lambda_1}\left(\mathbf{r}_B,\mathbf{r}_A,\text{i}\xi  \right) \right].
\label{Uchiral}
 \end{eqnarray}
 Resonant terms in the molecular polarisabilities have been avoided by means of infinitesimal imaginary shifts.

The interaction is obviously duality invariant, or in other words symmetric  over a simultaneous global exchange of electric and magnetic properties \cite{duality}.
In particular $U_{\lambda _1\lambda _2\lambda _3\lambda _4}$ can be obtained from $U_{\bar \lambda _1 \bar \lambda_2 \bar \lambda_3 \bar \lambda _4}$ ($\bar \lambda=\delta_{\lambda0}$) by using a duality transformation \cite{duality}.

Restricting our consideration to non-chiral molecules reduces Eq.~(\ref{Uchiral}) to the known results in Ref.~\cite{safari2}, which correspond to the terms with $\lambda_1=\lambda_2$ and $\lambda_3=\lambda_4$. Doing so, the abandoned terms are indeed the contributions from the chirality of one or both molecules which are the focus of this paper.

The purely electric contribution $U^{EE}$ and the chiral contribution $U^{CE}$ are stated in the main text. The chiral contribution $U^{CC}$ reads:
\begin{multline}
U^{CC}\left( \mathbf{r}_A,\mathbf{r}_B \right) =  - \frac{\hbar \mu _0^2}{\pi }\int\limits_0^\infty  \text{d}\xi \xi ^2 \text{Tr}\left\{ \chi _A^{em}\left( \text{i}\xi  \right) \right. \\
  \cdot \vec \nabla _A \times \tens{G}\left( \mathbf{r}_A,\mathbf{r}_B,\text{i}\xi  \right)   \times \overleftarrow \nabla_B \cdot \chi _B^{me}\left( \text{i}\xi  \right) \cdot \tens{G}\left( \mathbf{r}_B,\mathbf{r}_A,\text{i}\xi  \right) \\ 
 + \chi _A^{em}\left( \text{i}\xi  \right) \cdot \vec \nabla _A \times \tens{G}\left( \mathbf{r}_A,\mathbf{r}_B,\text{i}\xi  \right)\\
\left.  \cdot \chi _B^{em}\left( \text{i}\xi  \right) \cdot \vec \nabla _B \times \tens{G}\left( \mathbf{r}_B,\mathbf{r}_A,\text{i}\xi  \right) \right\}.
\end{multline}

\paragraph*{\label{Sec2}Scattering Green's tensor.
}
The scattering Green's tensor for a a chiral plate is introduced in the main text.
In the non-retarded limit the parallel and orthogonal components of the wave-vector coincide,  leading to the following scattering Green's tensor :
\begin{widetext}
\begin{equation}
\tens{G}^1_{\text{nr}}\left( \mathbf{r}_A,\mathbf{r}_B,\text{i}\xi  \right) =  \pm \frac{c}{4\pi \xi r_ + ^3}
\left( \begin{array}{ccc}
\frac{2xy\left( r_ + ^2\left( 3z_ +  - 2r_ +  \right) - z_ + ^3 \right)}{\left( x^2 + y^2 \right)^2} & \frac{\left( x^2 - y^2 \right)\left( r_ + ^2\left( 2r_ +  - 3z_ +  \right) + z_ + ^3 \right)}{\left( x^2 + y^2 \right)^2} & y \\
\frac{\left( x^2 - y^2 \right)\left( r_ + ^2\left( 2r_ +  - 3z_ +  \right) + z_ + ^3 \right)}{\left( x^2 + y^2 \right)^2} & \frac{2xy\left( r_ + ^2\left( 2r_ +  - 3z_ +  \right) + z_ + ^3 \right)}{\left( x^2 + y^2 \right)^2} & -x \\
-y & x & 0 \end{array} \right)
\end{equation}
\end{widetext}
where the positive sign is for a positive chiral plate, and the negative sign for a negative chiral plate. Furthermore $r_+^2=x^2+y^2+z_+^2$.

In general the non-retarded limit does not commute with the curl operation. We must take the curl of the full Green's tensor and only after make the non-retarded limit.  The action of the curl is equivalent to the action of $\text{i} \textbf{k}_+ \times$ on the left side, where $\textbf{k}_+=(k^\parallel \cos \varphi, k^\parallel \sin \varphi, \text{i}\kappa ^ \bot)$.  Taking into account $\text{i}\mathbf{k}_ +  \times \mathbf{e}_{s + } = \frac{\xi }{c}\mathbf{e}_{p + }$, $\text{i}\mathbf{k}_ +  \times \mathbf{e}_{p + } = -\frac{\xi }{c}\mathbf{e}_{s + }$ the curl of the Green's tensor reads:
\begin{multline}
\nabla_A \times \tens{G}^1\left( \mathbf{r}_A,\mathbf{r}_B,\text{i}\xi  \right) = \\
\frac{\xi}{8\pi ^2 c}\int\limits_0^{2\pi } \text{d}\varphi  \int\limits_0^\infty  \text{d} k^\parallel \frac{k^\parallel }{\kappa ^ \bot }\text{e}^{\text{i}k^\parallel \left( x\cos \varphi  + y\sin \varphi  \right)}\text{e}^{ - \kappa ^ \bot z_ + }\\
\times \left(  \mathbf{e}_{p + }\mathbf{e}_{p - }r_{p \to s}-\mathbf{e}_{s + }\mathbf{e}_{s - }r_{s \to p}  \right).
\end{multline}
In the non-retarded limit:
\begin{multline}
\left[\nabla_A \times \tens{G}^1\left( \mathbf{r}_A,\mathbf{r}_B,\text{i}\xi  \right)    \right]_{\text{nr}}=\pm \frac{c}{4\pi \xi r_ + ^5}\\
\times \left( \begin{array}{ccc}
2x^2-y^2-z_+^2 & 3xy & -3xz_+ \\
3xy & -x^2+2y^2-z_+^2 & -3yz_+ \\
3xz_+ & 3yz_+ & x^2+y^2-2z_+^2 \end{array} \right).\\
\end{multline}

\end{document}